\documentclass[aps,prl,showpacs,twocolumn,superscriptaddress,floatfix,a4paper]{revtex4-1}  % for review and submission
\usepackage{graphicx}  % needed for figures
\usepackage{dcolumn}   % needed for some tables
\usepackage{bm}        % for math
\usepackage{amssymb}   % for math
\usepackage{slashed}				%
\usepackage{color}				%
\usepackage{svgcolor}				%
\usepackage{comment}				%
\usepackage{amsmath}				%
\usepackage{multirow}				%
\usepackage{subfigure}				%
\usepackage{hyperref}				%
\begin{document}

% The following information is for internal review, please remove them for submission
\widetext
%\leftline{Version 1.3 as of \today}
%\leftline{Primary authors: Belyaev, Hall, King, Svantesson}
%\leftline{To be submitted to PRL}

\title{
%Distinguishing SUSY instances via gluino cascade decays:
%E6SSM versus MSSM.
Novel gluino cascade decays in $E_6$ inspired models 
}
	
\author{Alexander Belyaev}
\email[E-mail: ]{a.belyaev@soton.ac.uk}
\affiliation{School of Physics \& Astronomy, University of Southampton,
        Highfield, Southampton SO17 1BJ, UK}
\affiliation{Particle Physics Department, Rutherford Appleton Laboratory, 
       Chilton, Didcot, Oxon OX11 0QX, UK}
       
       \author{Jonathan~P.~Hall}
%\footnote{E-mail: \texttt{jonathan.hall@soton.ac.uk}}}
\email[E-mail: ]{jonathan.hall@soton.ac.uk}
\affiliation{School of Physics \& Astronomy, University of Southampton,
        Highfield, Southampton SO17 1BJ, UK}
	
\author{Stephen~F.~King}
%\footnote{E-mail: \texttt{king@soton.ac.uk}}}
\email[E-mail: ]{king@soton.ac.uk}
\affiliation{School of Physics \& Astronomy, University of Southampton, 
        Highfield, Southampton SO17 1BJ, UK}

\author{Patrik Svantesson}
\email[E-mail: ]{p.svantesson@soton.ac.uk}
\affiliation{School of Physics \& Astronomy, University of Southampton,
        Highfield, Southampton SO17 1BJ, UK}
\affiliation{Particle Physics Department, Rutherford Appleton Laboratory, 
       Chilton, Didcot, Oxon OX11 0QX, UK}

\date{\today}

\begin{abstract}
%An article usually includes an abstract, a concise summary of the work
%covered at length in the main body of the article. It is used for
%secondary publications and for information retrieval purposes.
%For PRL, the rule of thumb is that the abstract should be less than
%8 lines and the text (excluding authors, abstract but including tables,
%figures and references) should be less than 4 pages (leave about 20 lines
%empty on page 4) in two-column format.
%PRL and PRD papers have to have PACS (Phsyics and Astronomy Classification
%Scheme) numbers. Please see {\tt http://www.aip.org/pacs/} for the numbers
%relevant to your paper. A set of standard references can be found at the
%end of this example paper.
We point out that the extra neutralinos and charginos, generically appearing in a large class of $E_6$ inspired models, lead to distinctive signatures from gluino cascade decays involving longer decay chains, more visible transverse energy,  softer jets and leptons and less missing transverse energy than in the
Minimal Supersymmetric Standard Model (MSSM). On the one hand, this makes the gluino harder to discover for certain types of conventional analysis. On the other hand, the $E_6$ inspired models have enhanced 3- and 4-lepton signatures, as compared to the MSSM, making the gluino more visible in these channels.  After extensive
scans over the parameter space, we focus on representative
benchmark points for the two models, and perform a detailed Monte Carlo analysis of the 3-lepton channel,
showing that 
$E_6$ inspired models are clearly distinguishable from the MSSM in gluino cascade decays.

%The results suggest that, in such $E_6$ inspired models, the current search strategies and exclusion limits for the gluino at the Large Hadron Collider will have to be reconsidered. We propose that the focus of searches, in these cases, should be moved to other channels for which the discovery prospects of these models are optimal.

\end{abstract}

\maketitle

\section{Introduction}

The general idea of softly broken Supersymmetry (SUSY) provides a very attractive framework for physics models beyond the Standard Model (SM), in which the hierarchy problem is solved \cite{Chung:2003fi}.  In addition SUSY may provide a dark matter (DM) candidate, a route towards a grand unified theory (GUT), and indirect support for String Theory. However, it is not strictly necessary that SUSY provides all of the observed relic abundance or that the theory is perturbative right up to the GUT scale.

The simplest SUSY extension to the SM, the Minimal Supersymetric Standard Model (MSSM), has been subjected to particularly close scrutiny at the CERN Large Hadron Collider (LHC).
For example, searches at ATLAS \cite{Aad:2011ib} and CMS \cite{Chatrchyan:2011zy}, under certain assumptions, constrain the MSSM gluino mass to be greater than about 500--700 GeV, while the squarks of the first and second family are typically constrained to be heavier than about 1 TeV.
%although the
%third family squark masses may be well below 1 TeV, since current searches are not sensitive to their decays.
%This is significant since the third family squarks must be lighter than 1 TeV in order to provide a satisfactory
%solution to the hierarchy problem.
ATLAS and CMS indications of a Higgs boson with a mass in the region
$\sim124$--126~GeV \cite{AtlasTalk,CMSTalk} also suggest that
some of the third family squark masses should also exceed 1 TeV,
although this is not necessary in more general SUSY
models such as the Next-to-Minimal Supersymetric Standard Model (NMSSM)
\cite{Hall:2011aa,Arvanitaki:2011ck,King:2012is,Ellwanger:2009dp}.

The idea of SUSY is more general than either the MSSM or the NMSSM.
Here we shall focus on theoretically plausable 
models which are motivated by an $E_6$ gauge group, typically arising in
many string constructions, 
%a gauged $U(1)$ is a remnant of an $E_6$ gauge group, 
%and all anomalies are elegantly cancelled by 
and which involve three complete 27 representational families of $E_6$ at the TeV scale,
each family consisting of a complete 16-component SM family
$Q,u^c,d^c,L,e^c,\nu^c$, colour triplet and antitriplet $D,\overline{D}$, 
two Higgs doublets $H_u,H_d$ and one SM singlet $S$.
%However, if the right-handed neutrinos are neutral,
%they may acquire superheavy masses as in the standard see-saw mechanism.
%This special case is called the E$_6$SSM \cite{Ref1}, and the extra Abelian
%gauge group under which right-handed neutrinos are neutral is denoted $U(1)_N$.
For definiteness we shall consider a particular model called 
the E$_6$SSM \cite{Ref1}, where the right-handed neutrinos $\nu^c$ are decoupled at a high scale, 
although analogous results apply more generally to 
the other $E_6$ models \cite{Hewett:1988xc} with a low energy gauged $U(1)'$ \cite{Langacker:2008yv}. 
%We remark that, since $E_6$ arises in many string constructions, the observation of such a %$Z'$ boson and 
%TeV scale matter could also be viewed as providing indirect support for string theory.
It is well known \cite{neutralinodm} that in such $E_6$ models 
the neutralino and chargino sectors of the MSSM are augmented
by eight additional neutralino states: 
two associated with the fermionic partners to the $Z'$ and third singlet $S_3$
(denoted USSM neutralinos), two further singlinos $S_{1,2}$ and four neutral Higgsinos. 
It is the singlet $S_3$ which acquire a vacuum expectation value
({\sc vev}) $\langle S_3 \rangle =s/\sqrt{2}$ and provides an effective $\mu$-term, $\frac{\lambda s}{\sqrt{2}}H_{3u}H_{3d}$.
There are also two additional chargino states.
{Moreover, it has been pointed out that two of the extra singlet-dominated 
neutralino states {\em must} be light, with one of them most likely being the
lightest supersymmetric particle (LSP), providing a new possible source of dark matter \cite{neutralinodm} as well as interesting Higgs \cite{novelhiggs} phenomenology. 

In this paper we show that 
the augmented neutralino and chargino sectors of the $E_6$ models
lead to distinctive signatures from gluino decays.
This is significant since colour octet gluinos
%harder to find using current methods.
are expected to be pair produced at the LHC with rather large cross-sections,
%, providing 
%one of the leading candidates for the first new physics that can be discovered at the LHC.
%Gluinos are pair produced at the LHC, 
then, due to R-parity conservation, 
decay into pairs of the LSP together
with leptons and jets. We point out that in $E_6$ models, due to the augmented neutralino and chargino sectors,
the gluino cascade decays will 
generically involve more links in the decay chains 
and less missing transverse energy than in the MSSM.
To illustrate this, we perform scans over parameter space and focus on particular
benchmark points in the MSSM and E$_6$SSM which we subject to a detailed Monte Carlo (MC)
analysis. 
%For example, in the MSSM, the typical decay chain of the gluino will involve only one to three steps, 
%while in the E$_6$SSM, this is increased to three to five steps.
%Moreover, since the mass splitting between the lightest and the second lightest
%neutralinos is typically small, this will result in a higher multiplicity of 
%softer leptons and jets, providing challenging signatures.
%The results suggest that,  
%in such models, the current search 
%strategies  for MSSM should be reconsidered and adjusted for the E$_6$SSM case
%which would eventually lead to different exclusion limits for the gluino mass
%at the Large Hadron Collider.

%
\section{Parameter spaces}
WMAP \cite{WMAP} puts a bound on the LSP's relic density and the recent XENON100\, experiment \cite{xenon100} puts a bound on the direct detection cross-section for the LSP for a given relic density. These constraints exclude large portions of the parameter space for SUSY models. We have used \texttt{CalcHEP} \cite{calchep} and \texttt{MicrOMEGAs} \cite{micromegas} when scanning the parameter spaces of the MSSM and E$_6$SSM to pick out benchmarks which satisfy these constraints on the LSP as well as constraints from collider experiments. The scanning regions are presented in Tab.\ \ref{tab:scan} and points, including benchmarks, are plotted in the plane of the LSP relic density and the direct detection cross-section in Fig.~\ref{fig:scan}.

% {\color{magenta} Note that although E$_6$SSM neutralinos could constitute the whole of the observed dark matter relic density this would be in conflict with direct detection searches.
%This is because in the E$_6$SSM in order for the LSP to be heavy enough to annihilate efficiently at the time of thermal freeze-out it needs to contain significant inert Higgsino components and therefore has a rather large coupling to the SM-like Higgs and therefore a rather large direct detection cross-section \cite{novelhiggs}.
%Lower relic densities, as assumed here, are, however, consistent. }

The gluino decay chain length, $l$, relevant here is defined as the number of steps in the gluino decay chain after the first squark and illustrated in Fig.\ \ref{fig:length}. To be able to compare the models on a common basis the gaugino masses were fixed so that a gluino mass of 700 GeV, a wino mass around 300 GeV and a bino mass around 150 GeV were provided. {By contrast the squarks are all assumed to be much heavier, which is a characteristic feature of the constrained E$_6$SSM \cite{Athron:2009bs}, with universal high energy soft scalar and gaugino masses, although here we only consider the low energy version of the model. }

\begin{table}[t!]
	\centering
	\begin{tabular}{cccccc}
				&	\multicolumn{2}{c}{MSSM}	&	\multicolumn{2}{c}{E$_6$SSM}&	\\	
	parameter		&	min	&	max	&	min	&	max	&	\\
	\hline
	\hline
	$\tan\beta$		&	2	&	60	&	1.4	&	2	&	\\
	$|\lambda|$		&	-	&	-	&	0.3	&	0.7	&	\\
	\hline
	$s$			&	-	&	-	&	3.7	&	8	&	\multirow{4}{*}{\rotatebox{-90}{\hspace{-3mm}[TeV]}}\\
	$\mu$			&	-2 	&	2 	&	-	&	-	&	\\
	$A$			&	-3 	&	3 	&	-3	&	3	&	\\
	$M_A$			&	0.1 	&	2	&	1	&	5	&	\\
	\hline
	\end{tabular}
	\caption{The scanning region, where $\tan\beta=\frac{\langle H_{3u}\rangle}{\langle H_{3d}\rangle}$ is the ratio of the third family Higgs {\sc vev}s, $M_A$ is the mass of the psuedoscalar Higgs boson and $A$ is the soft trilinear coupling.
	A common squark and slepton mass scale was fixed to $M_{S}=2$ TeV. For the E$_6$SSM a large number of Yukawa couplings were scanned over which is omitted from this table.}
	\label{tab:scan}
\end{table}
%%%%%%%%%%%%%%%%%%
\begin{figure}[t]
	\centering
	\includegraphics[width=\columnwidth]{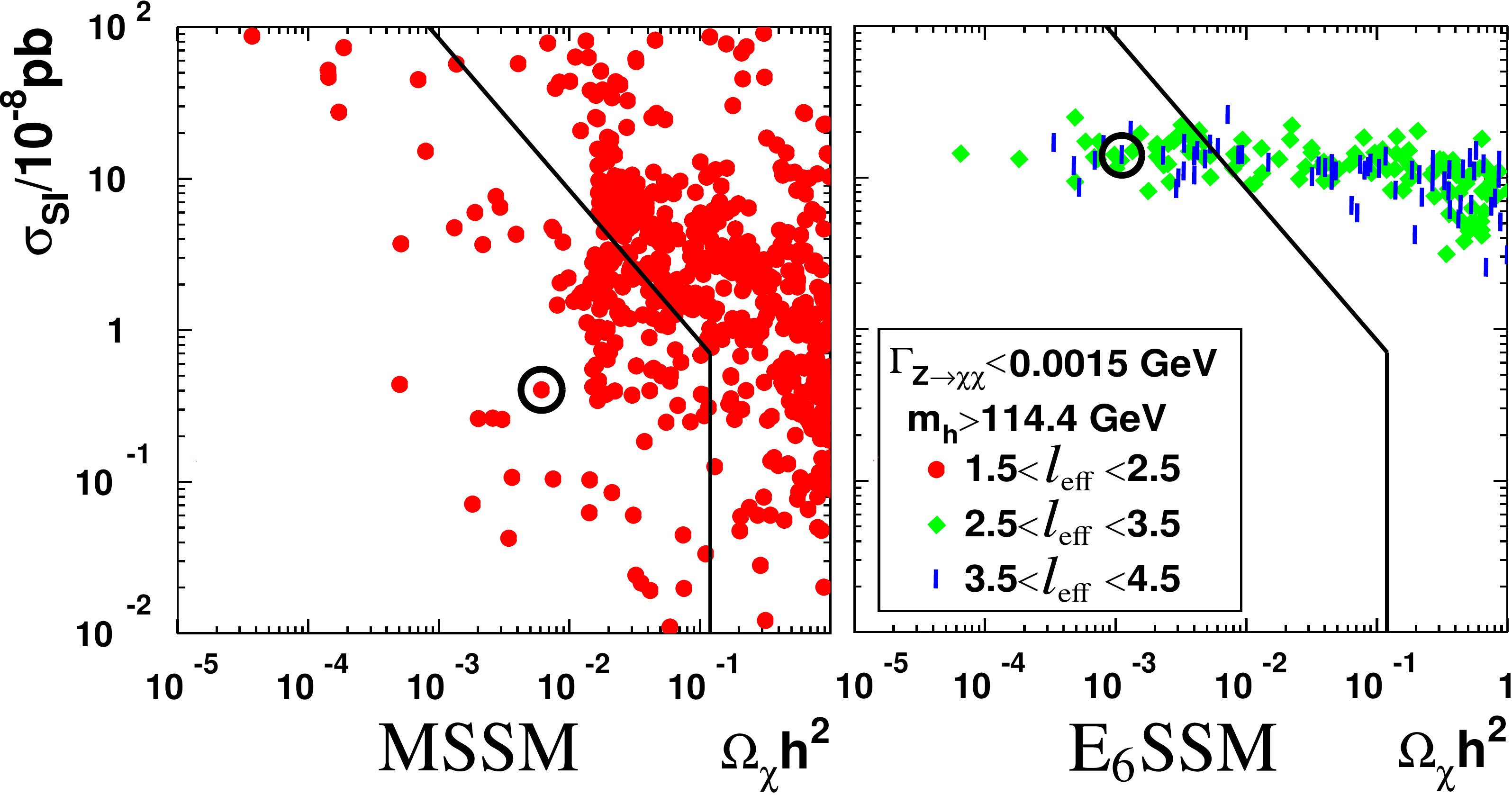}
	\caption{The scanned regions of the parameter spaces projected onto the plane spanned by the spin independent cross-section, $\sigma_{SI}$, and the relic density, $\Omega h^2$. {The area right of the vertical solid line is excluded by WMAP \cite{WMAP} and the area above the diagonal line is excluded by XENON100 \cite{xenon100} where the direct detection cross-section exclusion of the LSP gets weighted by its relic density.} The colouring represents the effective gluino decay chain length $l_{\mathrm{eff}}=\sum_l l\cdot P(l)$ for each point, where $P(l)$ is the probability for a chain length of $l$, as defined in Fig.\ \ref{fig:length}. The chosen benchmarks of MSSM and E$_6$SSM are encircled.}
	\label{fig:scan}
\end{figure}
\begin{figure}[ht]
	\centering
	\includegraphics[width=\columnwidth]{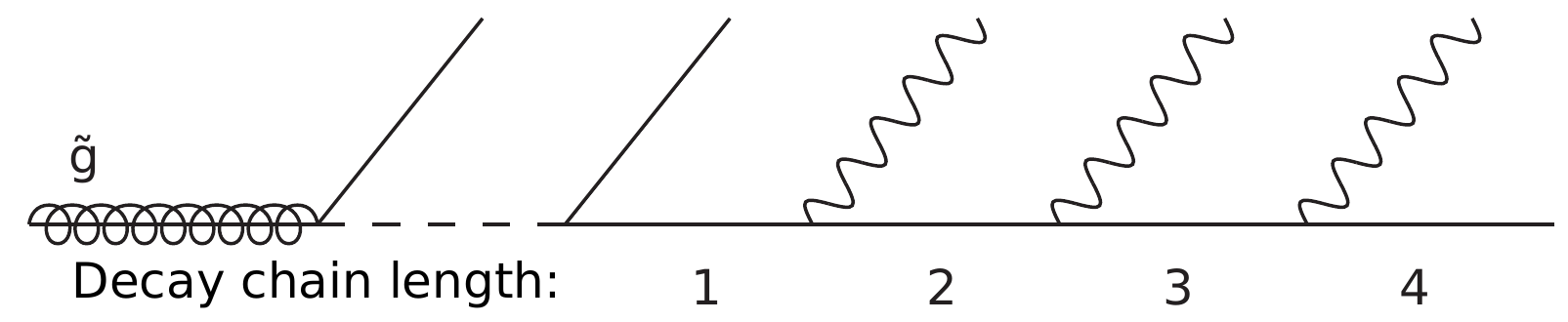}
	\caption{The definition of the gluino decay chain length, $l$.}
	\label{fig:length}
\end{figure}
%%%%%%%%%%%%%%%%%%
%
\section{Benchmarks}
The benchmarks considered for MC event analysis are summarized in Tab.\ \ref{tab:bm-table} and encircled in Fig.\ \ref{fig:scan}.
These benchmarks do not provide the whole observed amount of dark matter and so an additional source of dark matter is assumed.
% Other scenarios, where E$_6$SSM give the right amount of dark matter \cite{binodm}, have been considered in the same analysis \cite{gluinopheno}, but are not presented here.
{For the chosen MSSM benchmark
the LSP is bino-like and annihilates via the
pseudoscalar Higgs resonance, reducing its relic abundance below the observed value.}
{For the E$_6$SSM benchmark the overwhelmingly inert-neutralino-like LSP contains significant inert Higgsino components and 
annihilates via the Higgs resonance, near 125 GeV.}
\begin{table}
	%\centering
	\setlength{\tabcolsep}{.2mm}
	\begin{minipage}[t!]{.6\columnwidth}
	\hspace{-5mm}
	\begin{tabular}{ccccccccc}
				&		MSSM	&  E$_6$SSM	&\\
		\hline
		\hline
		$\tan\beta$	&		10	&	1.77	&\\
		$\lambda$	&		-	&	0.55	&\\
		\hline	
		$s$		&		-	&	5418	&\multirow{4}{*}{\rotatebox{-90}{[GeV]$\!\!\!\!\!\!\!\!\!\!\!\!\!\!\!\!\!\!\!\!\!\!\!\!\!\!\!\!\!\!\!\!\!\!\!\!\!\!\!\!\!\!\!\!\!\!\!\!\!\!\!$}}\\
		$\mu$		&		1578	&	(2107)	&\\
		$A$		&		-2900	&	-2200	&\\
		$M_A$		&		302.5	&	2736	&\\
		$M_1$		&		150	&	150	&\\
		$M_2$		&		285	&	300	&\\
		$M_{1'}$	&		-	&	151	&\\
		$m_{\tilde g}$	&		700	&	700	&\\
		\hline
		$P(l=1)$		&	0.231	&	$<10^{-9}$	& \\
		$P(l=2)$		&	0.769	&	$<10^{-4}$	&\\
		$P(l=3)$		&	0	&	0.287		&\\
		$P(l=4)$		&	0	&	0.782		& \\
		$P(l=5)$		&	0	&	0.005		& \\
		\hline
		$\Omega h^2$		&	0.00628	&	0.00114	&\\
		\hline
		$\sigma_{SI}$ 
					&	$0.04\times10^{-8}$	&	$15.3\times10^{-8}$	& \rotatebox{-90}{$\!\!\!\!\!$[pb]}\\
		\hline
	\end{tabular}
	\end{minipage}
	\setlength{\tabcolsep}{.8mm}
	\begin{minipage}[T]{.37\columnwidth}
	\hspace{-5mm}
	\begin{tabular}{cccccccc}
				&		MSSM	&  E$_6$SSM	\\%&\\
		\hline
		\hline
		${\tilde \chi^0_{M1}}$&		148.7	&	148.6	\\%&\multirow{6}{*}{\rotatebox{-90}{[GeV]}}\\
		${\tilde \chi^0_{M2}}$&		302.2	&	294.8	\\%&\\
		${\tilde \chi^0_{M3}}$&		1582	&	1459	\\%&\\
		${\tilde \chi^0_{M4}}$&		1584	&	1468	\\%&\\
		${\tilde \chi^\pm_{M1}}$&	302.2	&	298.7	\\%&\\
		${\tilde \chi^\pm_{M2}}$&	1584	&	1440	\\%&\\
		\hline
		${\tilde \chi^0_{U1}}$&		-	&	1254	\\%&\multirow{2}{*}{\rotatebox{-90}{$\!\!\!\!\!$[GeV]}}\\
		${\tilde \chi^0_{U2}}$&		-	&	1420	\\%&\\
		\hline	
		${\tilde \chi^0_{E1}}$&		-	&	62.7	\\%&\multirow{8}{*}{\rotatebox{-90}{[GeV]}}\\
		${\tilde \chi^0_{E2}}$&		-	&	62,8	\\%&\\
		${\tilde \chi^0_{E3}}$&		-	&	119.9	\\%&\\
		${\tilde \chi^0_{E4}}$&		-	&	121.1	\\%&\\
		${\tilde \chi^0_{E5}}$&		-	&	183.1	\\%&\\
		${\tilde \chi^0_{E6}}$&		-	&	184.4	\\%&\\
		${\tilde \chi^\pm_{E1}}$&	-	&	109.8	\\%&\\
		${\tilde \chi^\pm_{E2}}$&	-	&	117.8	\\%&\\
		\hline
		${h}$			&	124.4	&	125.4	\\%&\\%\multirow{2}{*}{\rotatebox{-90}{$\!\!\!\!\!$[GeV]}}\\
		\hline
	\end{tabular}\rotatebox{-90}{{\hspace{-0.4cm}[GeV]}}
	\end{minipage}
	\caption{Properties of the benchmarks. To the left: input parameters, including the soft gaugino masses $M_1$, $M_2$, $M_{1^\prime}$ and the physical gluino mass $m_{\tilde g}$, probabilities, $P(l)$, of certain guino decay chain lengths, $l$, and relic density and spin independent direct detection cross-section of the LSP. To the right: masses of neutralinos (absolute values), charginos and the lightest Higgs. In the notation for neutralino and chargino states the subscript $M$ denotes a MSSM-like state, $U$ a USSM-like state and $E$ an E$_6$SSM-like state. This distinction is reasonable since these sectors are very weakly coupled. The following number orders the states by mass. The squark and slepton mass scale is set to 2 TeV. Additional Yukawa couplings for the E$_6$SSM benchmark is given in Tab. \ref{tab:e6-yukawa}.}
	\label{tab:bm-table}
\end{table}
%%%%%%%%%%%%%%%%%%%%%%%%%%%%
\begin{table}
	\centering
	\begin{tabular}{cc}
	%			&E$_6$SSM\\
	%	\hline
	%	\hline
%		$\lambda$	& $5.5\times10^{-1}$\\
		$\lambda_{22}$	& $-5\times10^{-4}$\\
		$\lambda_{21}$	& $4.2\times10^{-2}$\\
		$\lambda_{12}$	& $4.5\times10^{-2}$\\
		$\lambda_{11}$	& $1\times10^{-3}$\\
	%	\hline
	\end{tabular}\vrule
	\begin{tabular}{cc}
	%			&E$_6$SSM\\
	%	\hline
	%	\hline
		$f_{d22}$	& $1\times10^{-3}$\\
		$f_{d21}$	& $6.844\times10^{-1}$\\
		$f_{d12}$	& $6.5\times10^{-1}$\\
		$f_{d11}$	& $1\times10^{-3}$\\
	%	\hline
	\end{tabular}\vrule
	\begin{tabular}{cc}
	%			&E$_6$SSM\\
	%	\hline
	%	\hline
		$f_{u22}$	& $1\times10^{-3}$\\
		$f_{u21}$	& $6.7\times10^{-1}$\\
		$f_{u12}$	& $6.4\times10^{-1}$\\
		$f_{u11}$	& $1\times10^{-3}$\\
	%	$x_{n\alpha}=z_\alpha$	& $1\times10^{-3}$\\
	%	\hline
	\end{tabular}
	\caption{Non-zero Yukawa couplings in the E$_6$SSM benchmark. The couplings $\lambda_{ijk}$ come from the terms $\lambda_{ijk}S_iH_{dj}H_{uk}$ in the superpotential. Here $\lambda_{333}=\lambda$,
		 $\lambda_{3\alpha\beta}=\lambda_{\alpha\beta} $,
		 $\lambda_{\alpha3\beta}=f_{d\alpha\beta} $,
		 $\lambda_{\alpha\beta3}=f_{u\alpha\beta} $,
		 $\lambda_{33\alpha}=x_{d\alpha} $,
		 $\lambda_{3\alpha3}=x_{u\alpha} $, and
		 $\lambda_{\alpha33}=z_{\alpha} $.
		 The $x$ and $z$ couplings are all $1\times10^{-3}$ and $\lambda=5.5\times10^{-1}$.}
	\label{tab:e6-yukawa}
\end{table}
%%%%%%%%%%%%%%%%%%%%%%%%%%%%
\begin{figure}[ht]
	\flushleft
	MSSM:\\
	\includegraphics[height=.2\columnwidth]{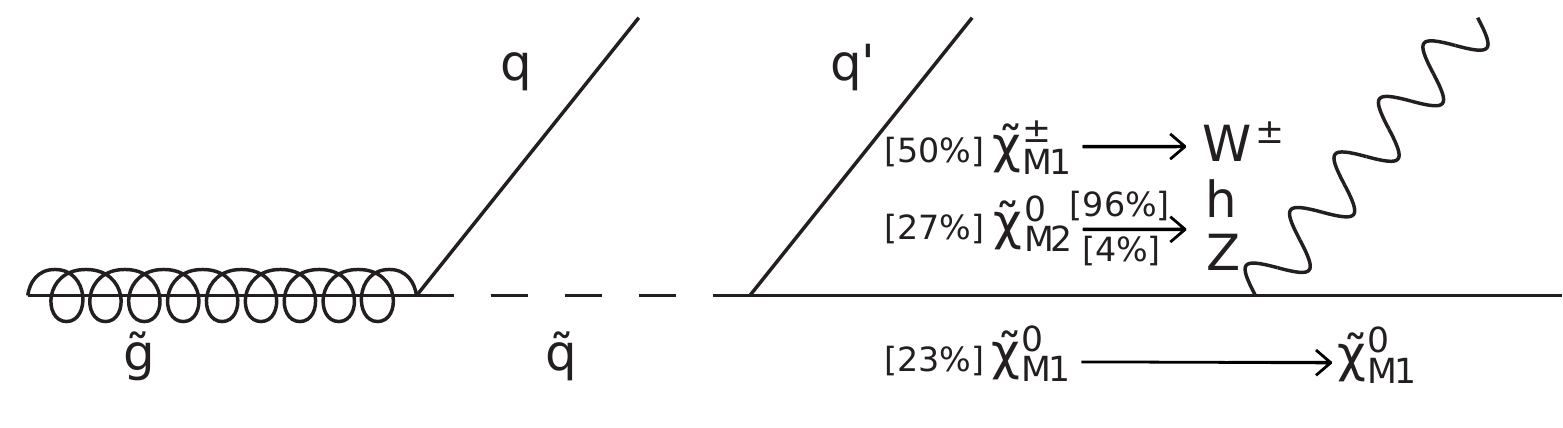}\\
	E$_6$SSM:\\
	\includegraphics[height=.2\columnwidth]{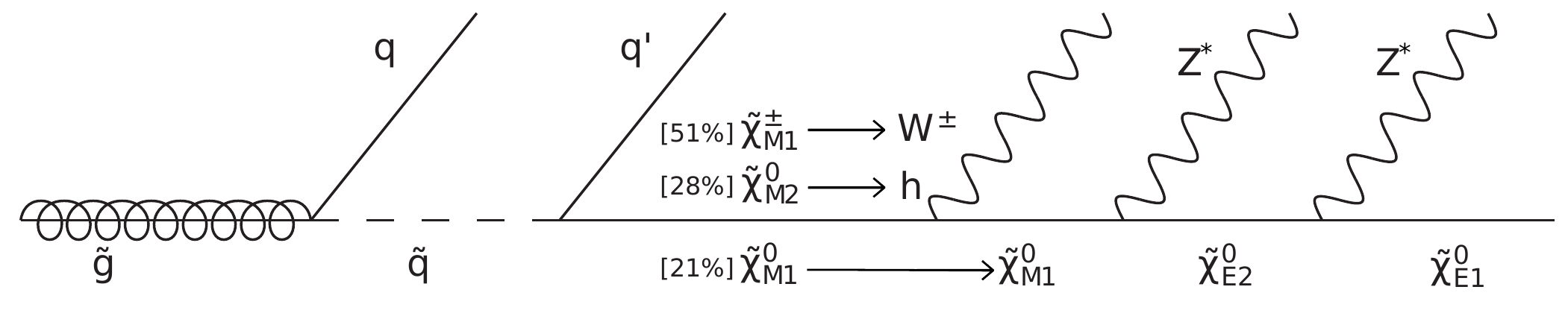}\\
	\caption{Feynman diagrams for the leading gluino decay chains for each benchmark. The branching ratios for produced particles are denoted in brackets.}
	\label{fig:diagrams}
\end{figure}
%%%%%%%%%%%%%%%%%%%%%%%%%%%
%
\section{Event analysis}
Since the E$_6$SSM introduces new neutralinos, naturally lig\-ht\-er than the MSSM LSP, the gluino decay chains will be longer than the MSSM's in general. This is confirmed and illustrated by the parameter scans in Fig.\ \ref{fig:scan} and benchmarks in Tab.\ \ref{tab:bm-table}. An effect of longer decay chains is that there will be less missing momentum in collider experiments. This affects the main SUSY searches based on jets and missing energy, e.g.\ \cite{Aad:2011ib} and \cite{cms}, which provide the best statistics and strongest exclusions. In these searches the E$_6$SSM is disfavoured compared to the MSSM and the acquired exclusions do not hold for this model. The main reason for the suppression of the E$_6$SSM comes from hard cuts on missing energy and its ratio with the effective mass. Our analysis show that the distributions for these variables are significantly different for these models. 
The distributions of the missing transverse momentum, $\slashed p_T$, and the effective mass, $M_{\mathrm{eff}}=\slashed p_T + \sum_{\mathrm{visible}}|p_T^{\mathrm{visible}}|$, before cuts are plotted in Fig.\ \ref{fig:ptmiss} for our benchmarks to illustrate this difference. 
{We have generated MC events using \texttt{CalcHEP} \cite{calchep} with CTEQ6L PDFs and we assume 30 fb$^{-1}$ of LHC data at $\sqrt{s}=8$ TeV. In this setup the cross-section for gluino pair-production at LHC for our benchmarks is 93 fb implying 2790 produced pairs of gluinos. }
Another important feature of the long decay chains of the E$_6$SSM is the increase in lepton as well as jet multiplicity, as shown in Fig.\ \ref{fig:multiplicity}. This feature allows us to rely on multi-lepton requirements for background reduction rather than cuts on missing energy. There is a significant loss of statistics by using this strategy, however, it turns out to be the most favourable channel of discovery and a channel in which instead the E$_6$SSM is largely dominant compared to the MSSM. 
In Fig. \ref{fig:ptmiss-3lep} the missing transverse momentum and the effective mass are plotted for our benchmarks and SM backgrounds after requiring three leptons ($\mu$ or $e$) with $p_T>10$ GeV,  $|\eta|<2.5$ and $\Delta R(\mathrm{lepton,jet})>0.5$ and where the leading lepton has $p_T>20$ GeV. 
We have also applied a  Gaussian smearing of  lepton and jet energies to take into account the detector energy resolution typical for the ATLAS and CMS detectors. % Gaussian smearing: Delta=0.5 for jets and Delta=0.15 for leptons
The dominant background is coming from $ZWj$ and $t\bar t V$, other important contributions come from $ZW$ and $t\bar t$. {Our background predictions agree well with backgrounds used in multi-lepton searches by CMS \cite{CMS-PAS-SUS-11-013} and ATLAS \cite{ATLAS-CONF-2012-001}.}
By choosing a signal region defined by the cut $M_{\mathrm{eff}}>900$ GeV, $S=36$ signal events and $B=5$ background events are expected. Using the definition of statistical significance, $S_{12}=2(\sqrt{S+B}-\sqrt{B})$, valid for small statistics \cite{sig-cms,sig-bityukov}, a $8.3\sigma$ excess is predicted. 
Even for $15$fb$^{-1}$ integrated luminosity the signal  is discoverable with a $5.9\sigma$ excess.
%%%%%%%%%%%%%%%%%%%%%%%%%%%%%%
\begin{figure}[h!]
	\includegraphics[width=.49\columnwidth]{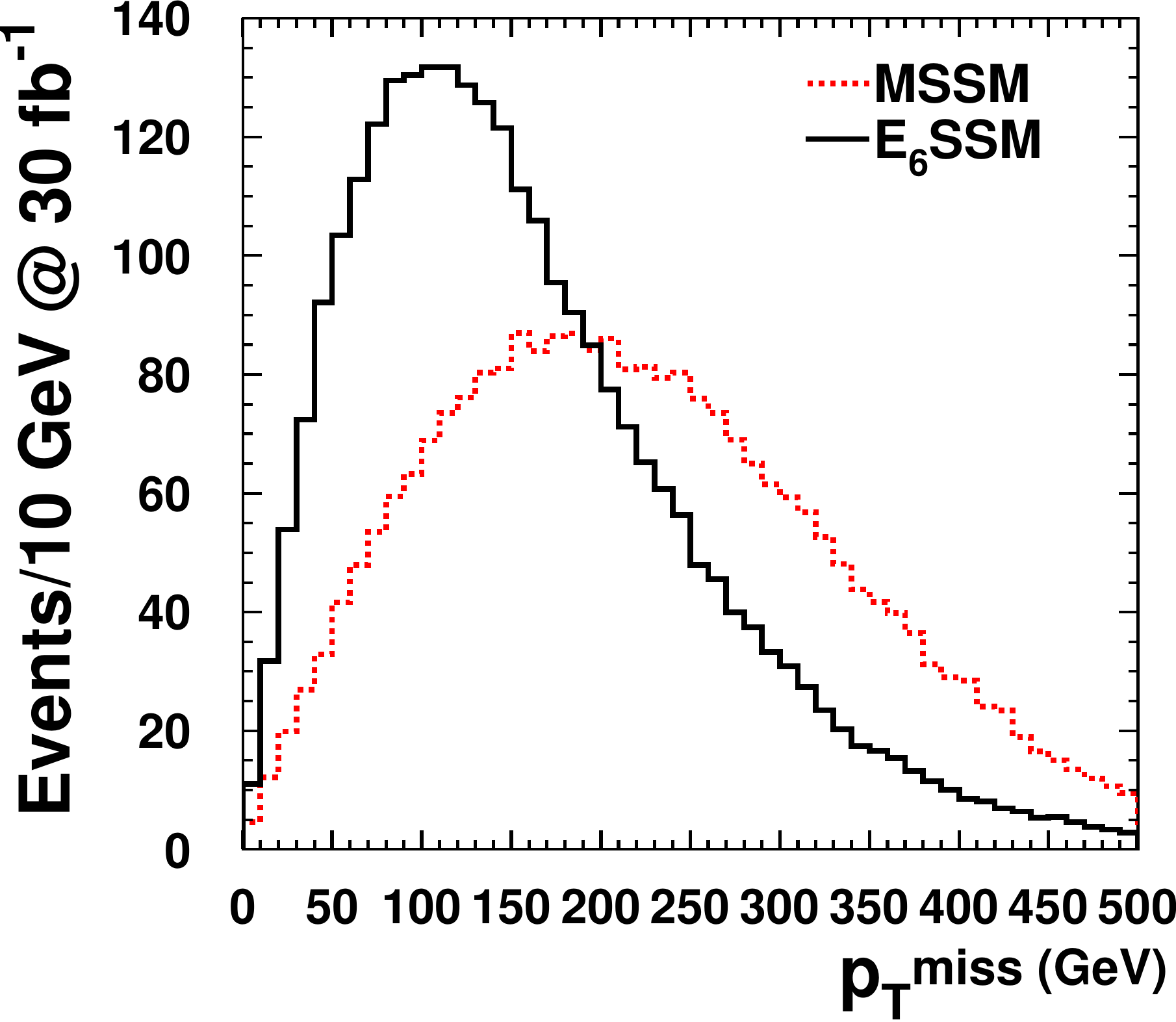}
	\includegraphics[width=.49\columnwidth]{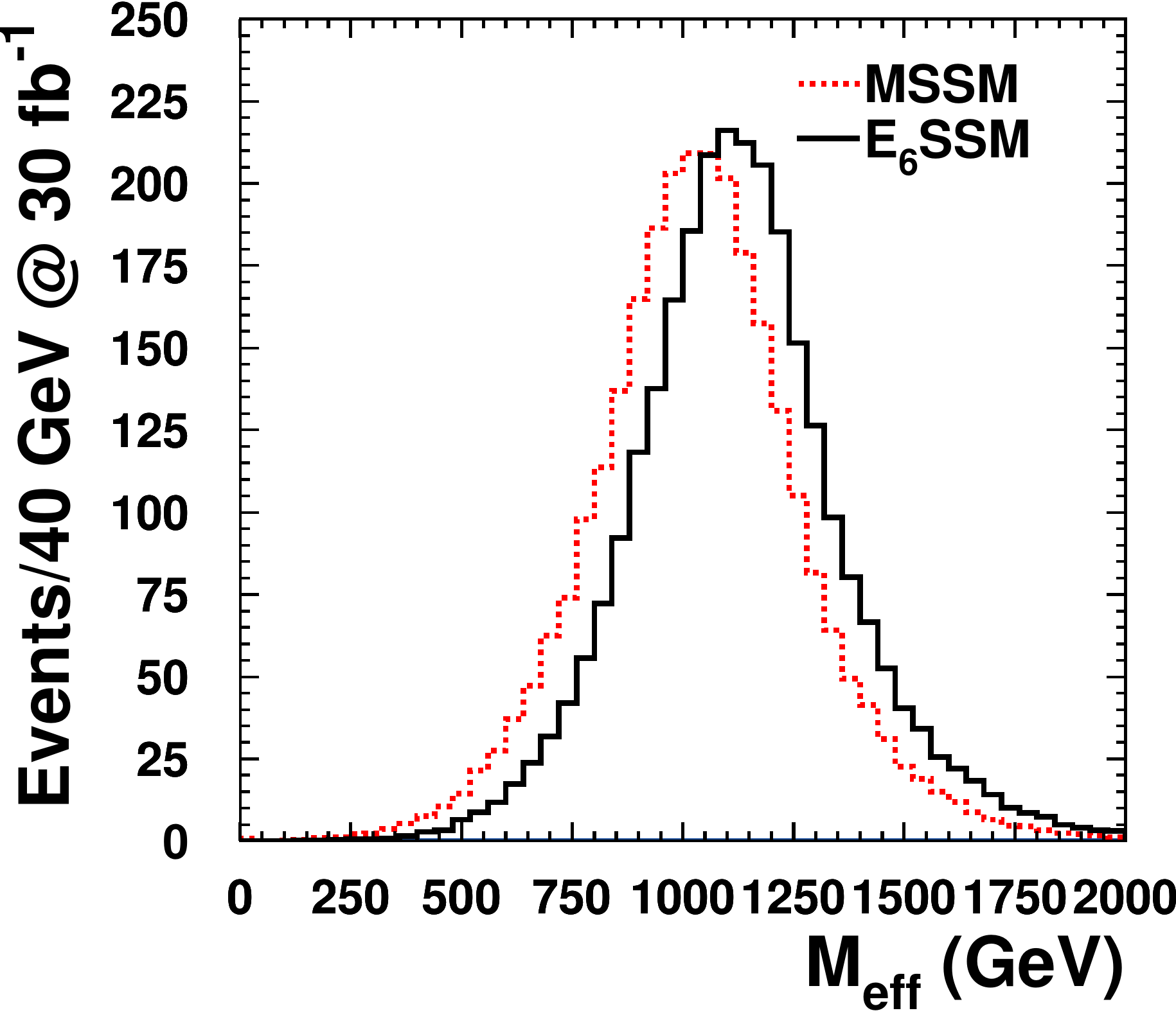}
	\caption{Missing transverse momentum and the effective mass before cuts.}
	\label{fig:ptmiss}
\end{figure}
%%%%%%%%%%%%%%%%%%%%%%%%%%%%%%
\begin{figure}[h!]
	\centering
	\includegraphics[width=.49\columnwidth]{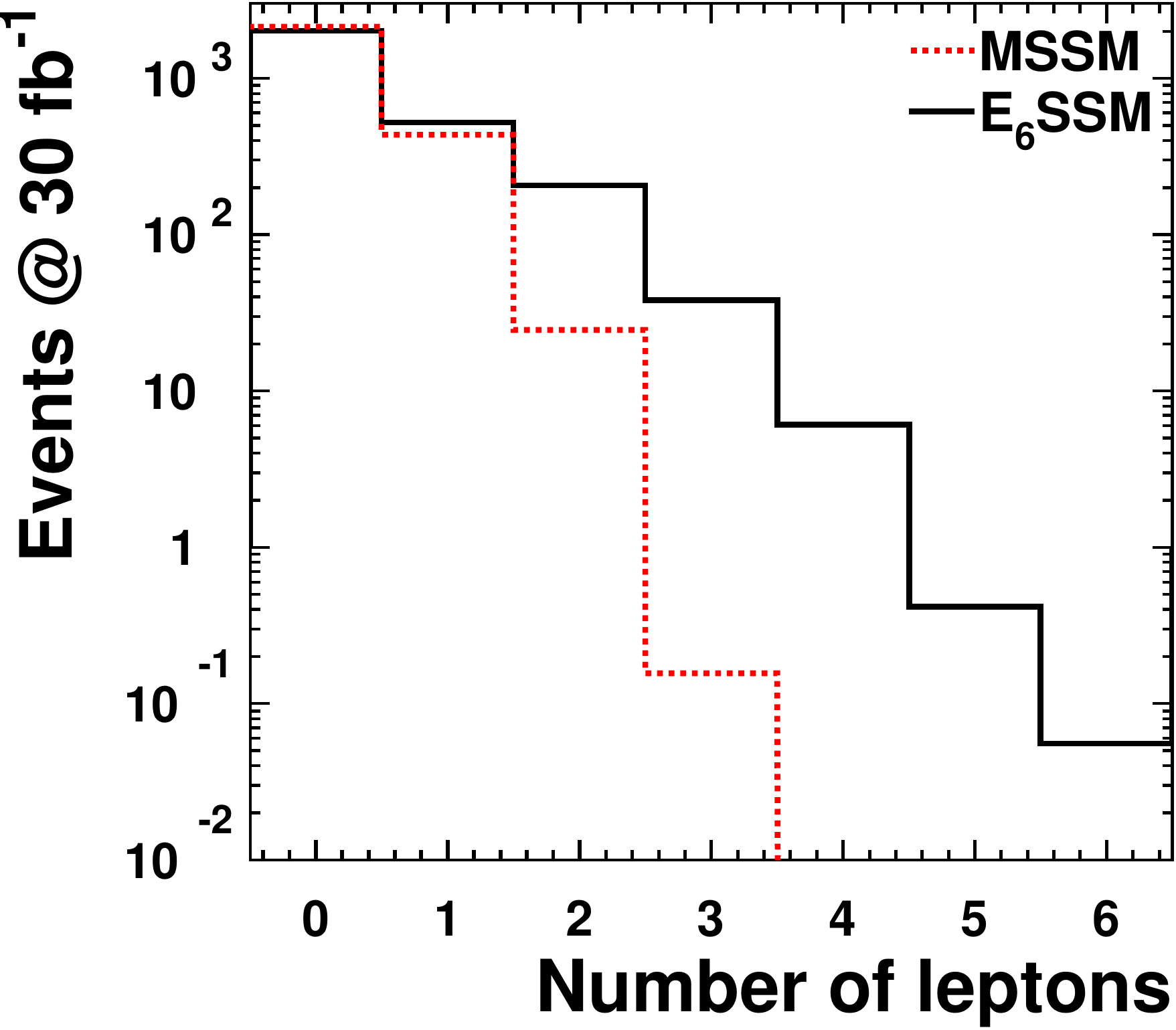}
	\includegraphics[width=.49\columnwidth]{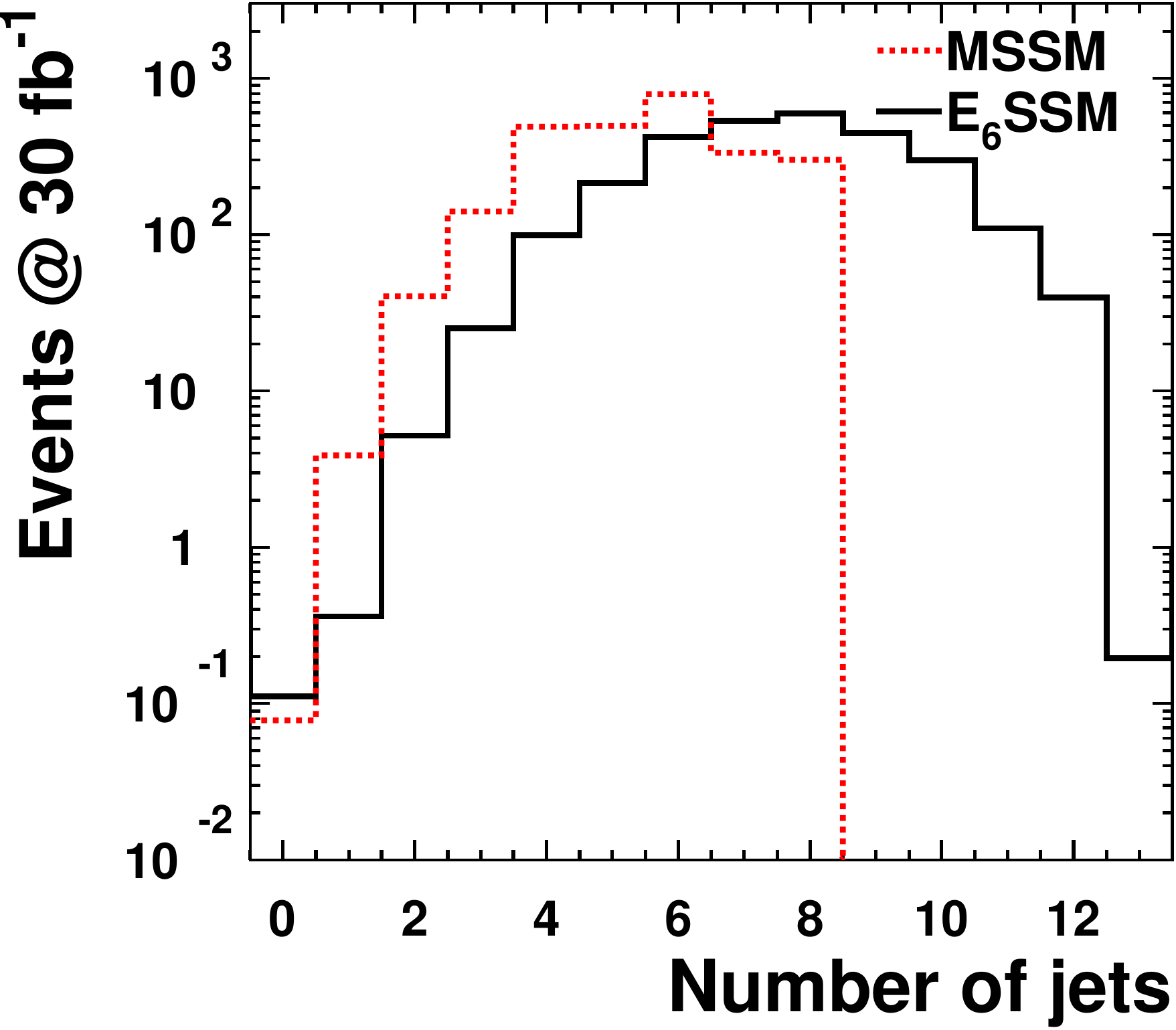}
	\caption{Lepton multiplicity and jet multiplicity, requiring $p_T>10$ GeV, $|\eta|<2.5$ and $\Delta R(\mathrm{lepton,jet})>0.5$ for leptons and $p_T>20$ GeV and $|\eta|<4.5$ for jets.}
	\label{fig:multiplicity}
\end{figure}
%%%%%%%%%%%%%%%%%%%%%%%%%%%%%%
\begin{figure}[h!]
	\includegraphics[width=.49\columnwidth]{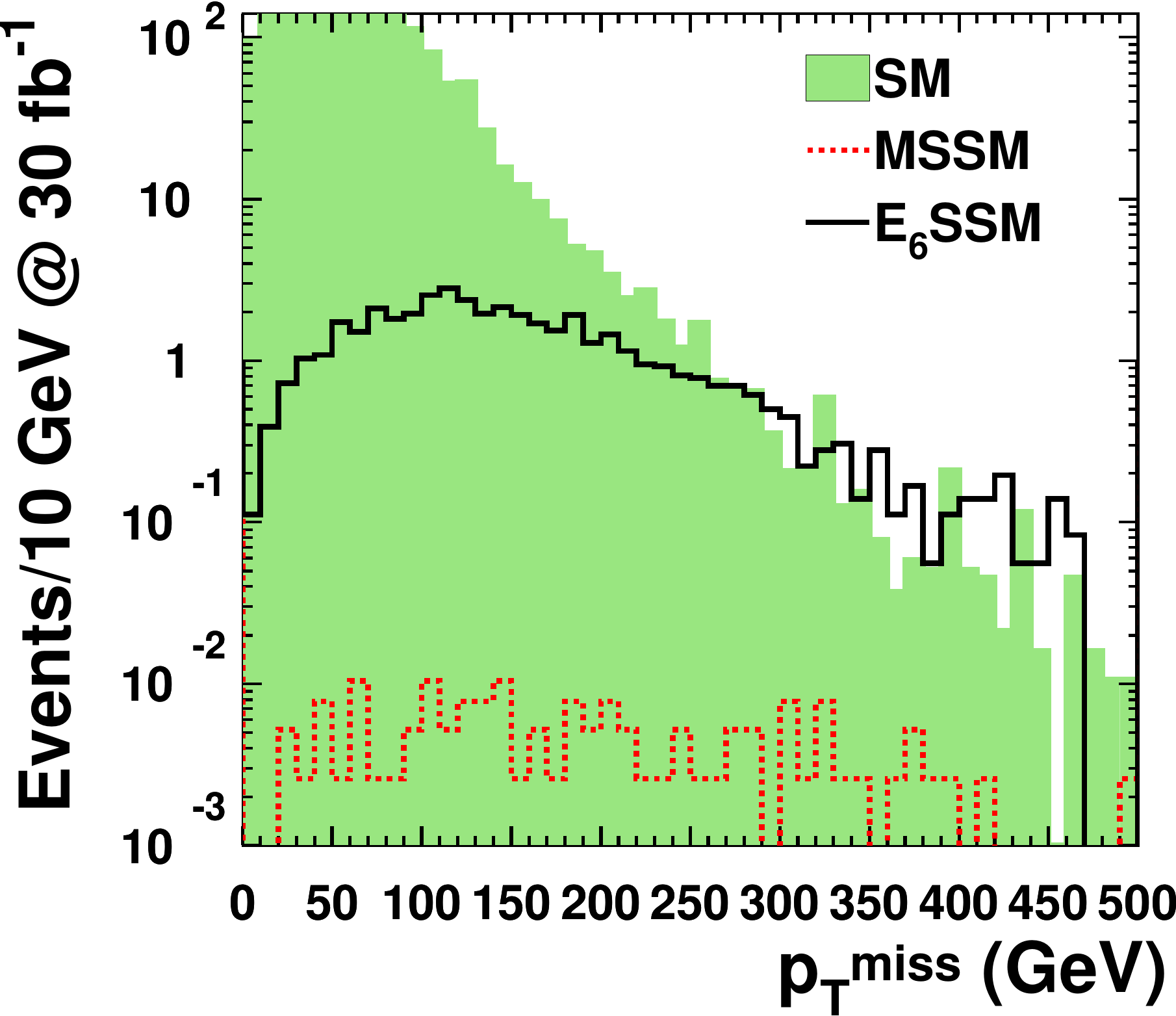}
	\includegraphics[width=.49\columnwidth]{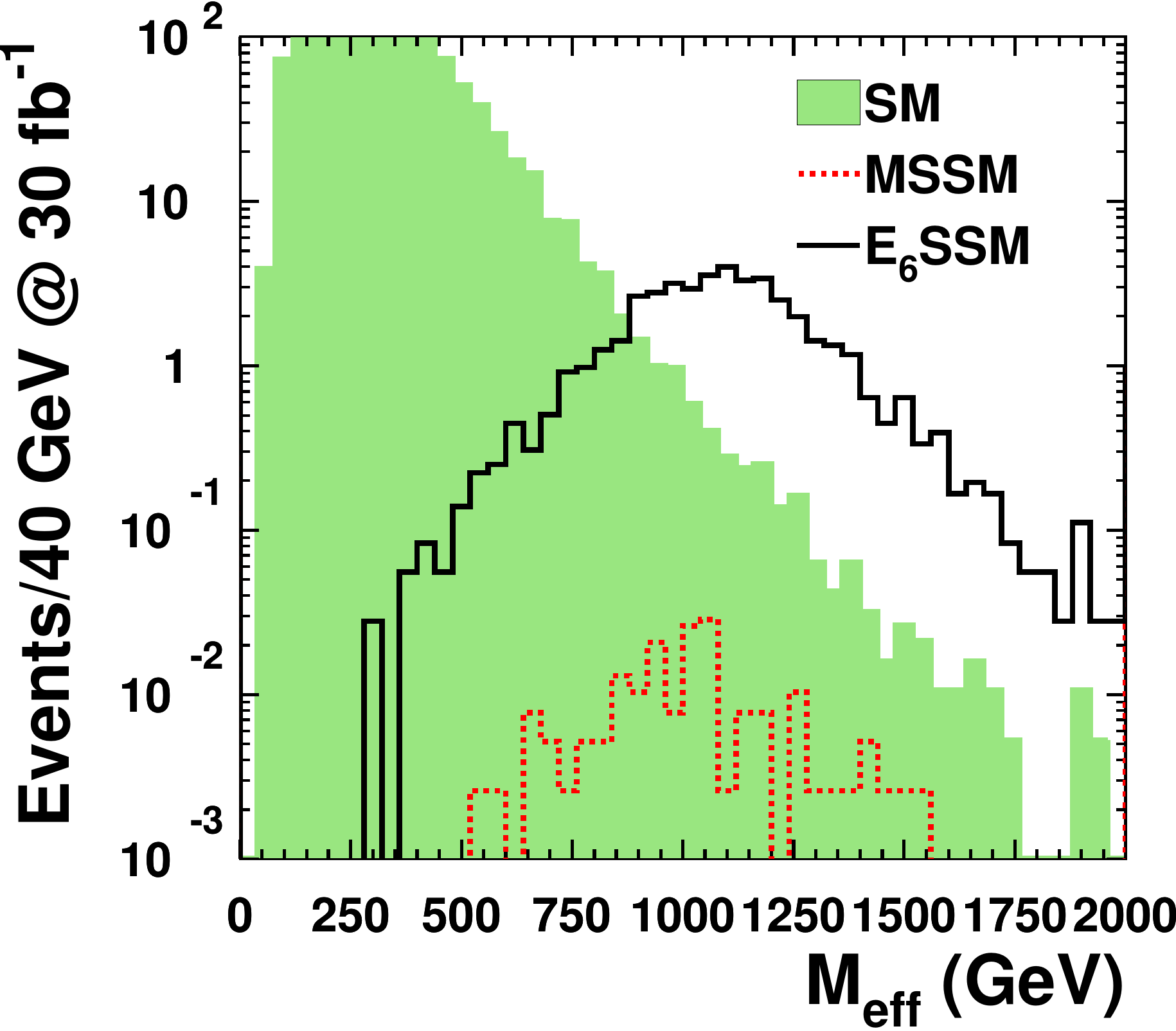}
	\caption{Missing transverse momentum and effective mass after requiring three leptons, one with $p_{T}>20$ GeV and two with $p_{T}>10$ GeV. When applying a cut on the missing transverse momentum an estimation of the deviation from the SM can be made. For $\slashed p_T>275$ GeV we expect $B=3.24$ SM events and $S=4.4$ events from the E$_6$SSM gluinos, which implies a $1.9\sigma$ excess. When instead applying a cut on the effective mass $M_{\mathrm{eff}}>900$ GeV we expect $S=36.4$ events from the E$_6$SSM and $B=5.0$ events from the SM, which implies a $8\sigma$ excess.
}
	\label{fig:ptmiss-3lep}
\end{figure}
%\newpage
%%%%%%%%%%%%%%%%%%%%%%%%%%%%%%
%%%%%%%%%%%%%%%%%%%
%%%%%%%%%%%%%%%%%%%
%%%%%%%%%%%%%%%%%%%
\section{Conclusions}
 
%Supersymmetry is an attractive candidate for new physics beyond the SM, but the idea
%is more general than the MSSM or NMSSM.
% and, due to its large cross-section,  the gluino may be the first superpartner to be discovered at the LHC. However, the decay signatures of the gluino may be very different depending on the SUSY model, and different decay channels may be favoured in different models.
%To illustrate this 
%we have considered here two very different SUSY models, namely the 
%We have considered novel gluino cascade decays in $E_6$ inspired models 
%MSSM and the E$_6$SSM which can lead to very different gluino decay signatures, requiring different gluino search 
%strategies in each case. This is due to 
We have  demonstrated that the extra neutralinos and charginos, generically appearing in $E_6$ inspired models,
lead to gluino decays involving longer decay chains, more visible transverse energy,  softer jets and leptons and less missing transverse energy than in the MSSM. This makes the $E_6$ inspired gluino harder to discover for certain types of conventional analysis based on large missing energy with a conventional multi-jet or 1-lepton analyses, and it is possible that the gluino signal 
in the E$_6$SSM could be missed in these channels. 
On the other hand, the $E_6$ signatures have an enriched lepton multiplicity, providing distinctive 3- and 4-lepton decay modes, which are disfavoured in the MSSM, making the $E_6$ inspired gluino more visible in these channels. 
%To illustrate these features we have performed scans over parameter space and have selected two particular
%benchmark points which we have subjected to a detailed Monte Carlo analysis.
%We have shown that indeed the gluino signatures in the E$_6$SSM have large visible and small missing $p_T$,
%with the effect of these features cancelling in $M_{\mathrm{eff}}$. In general, large jet and lepton multiplicity searches are favoured for $E_6$ inspired models and 
 For example, the 3-lepton signature from gluino cascade decays, may easily be visible for the 
E$_6$SSM, while the analogous signal for the MSSM
would be buried under the background.

In conclusion, the gluino is not only a key prediction of SUSY, but 
also distinguishes different SUSY models since 
its visibility depends strongly on the model and on the search channel.
These specific features of the E$_6$SSM and analogous models should be taken into account and included in current experimental searches. This could provide different limits for the E$_6$SSM gluino,
as compared to the MSSM, or perhaps even to an earlier SUSY discovery.

%%%%%%%%%%%%%%%%%%%
\section{Acknowledgements}
We would like to thank Alexander Pukhov for necessary improvements of \texttt{CalcHEP} and \texttt{MicrOMEGAs}. PS thanks the NExT institute and SEPnet for support. AB thanks the NExT Institute and Royal Society for partial financial support.
SFK acknowledges UNILHC 237920.

\end{document}